\documentstyle[12pt,aasms4]{article}
 
\slugcomment{To appear in the Astronomical Journal (accepted Nov. 26, 1997)}
 
\begin{document}

\title{A Sub-kpc Disk in MRK 231}
 
\author{C.L. Carilli, J.M. Wrobel, J.S. Ulvestad}
\affil{NRAO, P.O. Box O, Socorro, NM, 87801, USA \\}
\authoremail{ccarilli@nrao.edu}
 
\begin{abstract}

We present imaging with the Very Long Baseline Array of the  neutral
hydrogen 21cm absorption line system seen toward the nuclear regions
of Mrk 231 at z$_\odot$ = 0.04217, and imaging of the radio continuum
emission at 1.4 GHz on scales ranging from  a few parsecs to a few
hundred parsecs. These data  
indicate the existence of a sub-kpc gas disk in Mrk 231, as seen
in HI 21cm absorption and in radio continuum emission. 
The radio continuum morphology is consistent with a disk of
maximum radius of 440 mas  (260 h$^{-1}$ pc), at an
inclination angle of 45$^o$, with a major axis oriented east-west.
The HI 21cm absorption shows an east-west gradient in position and velocity
of about $\pm$ 110 km s$^{-1}$ out to radii of  100 mas (60
h$^{-1}$ pc). We identify this HI and radio continuum 
disk as the inner part of the molecular disk seen on a factor three
larger scale. The physical conditions for the thermal and
non-thermal gas in the sub-kpc disk of Mrk 231 are similar to those
proposed for compact nuclear starburst galaxies,
and in particular to the conditions proposed for the sub-kpc gas disk
in Arp 220. From the neutral hydrogen velocity field we
derive a gravitational mass enclosed within a 50 
h$^{-1}$ pc radius of 3$\times$10$^{8}$ h$^{-1}$
M$_\odot$, and from the radio continuum emission 
we derive a massive star formation  rate in the disk of 60
M$_\odot$ year$^{-1}$. 

We also present a search for HI 21cm absorption associated with the
optical broad absorption line (BAL) systems toward Mrk 231. 
We do not detect HI 21cm absorption associated with any of
the optical BAL systems. These negative results require  that 
the neutral atomic gas in the BAL clouds  be fairly warm
(T$_s$ $>$ 50 K), unless the  NaI abundance is higher than
solar, or the dust-to-gas ratio is higher than Galactic, or
the observed extinction toward the nucleus of Mrk 231 is not
due to the BAL gas.

\end{abstract}
 
\keywords{ galaxies:seyferts, active, ISM - quasars:absorption lines -
radio lines: galaxies} 

\section {Introduction}

The Seyfert 1 galaxy Mrk 231 has an infrared (IR) luminosity of
3x10$^{12}$ h$^{-2}$ L$_\odot$, making it the most luminous infrared
galaxy in the sample of  ultra-luminous infrared galaxies (ULIRGs)
of Sanders et al. (1988).\footnote{We use h $\equiv$ ${H_o}\over{100}$ km
s$^{-1}$ Mpc$^{-1}$, giving a scale of: 1 mas = 0.6 h$^{-1}$ pc at the
redshift of Mrk 231, z = 0.042.} Most of the infrared  emission
from Mrk 231 comes from a region less than 0.5$''$ (300 h$^{-1}$ pc)
in size (Matthews et al. 1987). Imaging of the CO emission from Mrk
231  by Bryant and Scoville (1996) shows a molecular disk on a scale
of about 1$''$. Deep optical imaging of Mrk 231 reveals a disturbed
parent galaxy, perhaps with a weak secondary nucleus, and 
with clear tidal features indicating a
merger within the last 10$^{9}$ years (Hamilton and Keel
1987, Hutchings and Neff 1987, Armus et al. 1994, Surace et al. 1997).
Radio continuum imaging shows an active radio
nucleus with a high surface brightness core plus radio lobes 
extending both north and south to about 30 mas (Ulvestad, Wrobel, and
Carilli 1997a).

Sanders etal. (1988) have proposed that ultraluminous infrared
galaxies are part of a sequence in which merging spiral galaxies
evolve through a starburst phase, eventually becoming  optical
QSOs (see also Sanders and Mirabel 1996). The early stage of the
merger (10$^7$ years) involves molecular cloud collisions, 
leading to a `funneling of the gas into the merger nucleus'. This gas
then fuels a  nuclear starburst on a timescale of about 10$^8$ years.
An important mechanism in this sequence entails
the `formation of a self-gravitating gas disk'
in the galaxy center on a scale $\le$  1 kpc. This disk would be
dissipative, driving gas condensation further into the
nucleus, and fueling the active nucleus. Radiation and/or winds from
the active nucleus  would eventually clear-out the obscuring
material, revealing an optical QSO. Scoville, Yun, and Bryant (1997)
show that the physical conditions in the
disk may be conducive to star formation. Sanders etal. propose that Mrk
231 represents a late stage in this process, namely the `dust
enshrouded QSO' stage. The origin of the massive black hole remains
uncertain (Weedman 1983). 

An important question concerning the Sanders et al. (1988) model
is: do sub-kpc gas disks exist in ULIRGs? 
Such gas disks have been proposed in Mrk 231 and Arp 220, based on 
molecular imaging observations (Bryant and Scoville
1996, Scoville, Yun, and Bryant 1997).
In this paper we present high resolution imaging of the neutral
hydrogen 21cm absorption line system seen toward the nuclear regions of Mrk
231 (Heckman, Balick 
and Sullivan 1978, Dickey 1982), and of the radio continuum emission
at 1.4 GHz on scales ranging from a few parsecs to a few hundred
parsecs. These data support the existence of a
sub-kpc gas disk in Mrk 231, as seen both in HI 21cm absorption
and in radio continuum emission. These data also suggest that star
formation activity may be significant in the disk.

We also present a search for HI 21cm absorption associated with the
optical broad absorption line systems toward Mrk 231. 
One of the more interesting observations of Mrk 231 has been the detection
of low ionization, time variable, broad absorption lines
systems (BAL) toward the optical nucleus, with out-flow velocities of
a few thousand km s$^{-1}$ 
(Boksenberg etal. 1978, Boroson et al. 1991, Kollatschny, Deitric, and
Hagen 1992, Forster et al. 1995). Boksenberg etal. (1978) suggest that
much of the dust giving rise to the A$_V$ = 2 toward the nucleus of
Mrk 231 could be associated with these low ionization BAL clouds. 
We do not detect HI 21cm absorption associated with any of
the BAL systems. These data provide interesting
constraints on the  physical conditions in the BAL gas seen toward Mrk 231. 

\section{Observations, Calibration, and Imaging}

\subsection{VLBA}

Observations of Mrk 231 were made on December 27, 1996 
with the Very Long Baseline Array (VLBA) operated by the National
Radio Astronomy Observatory (NRAO; Napier et al. 1994). 
These observations included the phased Very Large Array (VLA) as an
element in the very long baseline array. The VLA was in its largest (33 km)
configuration. The pass band was 
centered at the frequency of the neutral hydrogen 21cm line 
at a heliocentric redshift of: z =  0.04216, or cz = 12640 km
s$^{-1}$. The total bandwidth was 8 MHz, using two orthogonal
polarizations, 256 spectral channels, and 2 bit correlation. 

All data reduction was performed using the Astronomical Image
Processing System (AIPS) of the NRAO.
Standard {\sl  a priori} gain calibration was performed using the measured
gains and system temperatures of each antenna.
The compact radio source J1219+48 was observed every 20 minutes, and this
source was used to determine the fringe rates and delays during the
observations. These solutions were then applied to the Mrk 231 data.
The source 3C 345 (J1642+39) was used to calibrate the 
frequency dependent gains (band pass calibration). The 
source J1310+32 was used to check 
the absolute gain calibration, and to check the band pass calibration.
The results showed agreement of observed and expected flux densities to
within 10$\%$, and bandpass calibration accuracy to better than 2$\%$
(peak-to-peak residual variations).

After application of the delay and rate solutions, and band pass
calibration, a continuum data set for Mrk 231 was generated by
averaging off-line channels. This continuum data set was then used for
the hybrid imaging process, which involves iterative imaging and
self-calibration of the 
antenna-based complex gains (Walker 1985). The final iteration
involved both phase and amplitude calibration with a 1 minute
averaging time. The  self-calibration solutions were applied to 
the  spectral line data set. The spectral line data were then
analyzed at various spatial and spectral resolutions by 
tapering the visibility data, and by  smoothing in frequency. 
Standard analysis of the spectral line image cubes involved
subtracting the continuum emission by fitting linear spectral 
baselines at each position to the off-line channels using the AIPS
routine IMLIN, and then deconvolving the synthesized beam using the 
Clark `CLEAN' algorithm  as implemented in the AIPS routine APCLN
(Clark 1980). The channel images were CLEANed 
to a residual level equal to twice the  root mean square
(rms) noise level. 

\subsection{VLA}

Observations of Mrk 231 were made on January 27, 1993  with the VLA  in
the 10 km configuration (Napier, Thompson, and Ekers 1983),
centered on the frequency of the neutral hydrogen 21cm line at the
heliocentric redshifts of the three broad absorption line systems seen
toward Mrk 231: z = 0.01554 (cz = 4660 km s$^{-1}$), z = 0.02085 (cz =
6250 km s$^{-1}$), and z = 0.02668 (cz = 8000 km s$^{-1}$). One
polarization was received, and on-line Hanning smoothing was applied. 
Observations at 4660 km s$^{-1}$ and 6250 km s$^{-1}$ used 
63 spectral channels with a channel width of 97.6 kHz, while those at 8000
km s$^{-1}$ used 31 spectral channels with a channel width of 390.6
kHz. Absolute gain, complex gain, and band pass calibration were
performed on 3C 286. After application of the standard calibration
solutions, a continuum data set was generated by averaging all
channels, and this data set was used for iterative self-calibration of the
antenna based complex gains (all the channels were
used since no line signal was detected). 
The self-calibration solutions were then
applied to the spectral line data. The continuum emission was then
subtracted from the spectral line image cubes by fitting linear spectral 
baselines at each position to all the channels.

Observations of Mrk 231 were also made on August 31, 1997  with the VLA  in
the 3 km configuration,
centered on the frequency of the neutral hydrogen 21cm line at the
heliocentric redshift of 0.04212 (cz = 12640 km s$^{-1}$). Two orthogonal
polarizations were received, 
with 63 spectral channels, with a channel width of 48.8 kHz, and
on-line Hanning smoothing was applied. The data were calibrated,
self-calibrated, and imaged as described above.  We assume the errors
in absolute gain calibration are 3$\%$ for the VLA observations
(Carilli et al. 1991). 

\section{Results}

\subsection{Radio Continuum}

The VLA image of Mrk 231 at 1.4 GHz at 5$''$ resolution made from
data taken in January 1993 is shown in
Figure 1. The image shows a compact source with a peak
surface brightness of 240 $\pm$ 7 mJy beam$^{-1}$.
There is also low surface brightness emission extending at least 30$''$
to the south, and the total source flux density 
is 320 $\pm$ 9 mJy. A detailed discussion of the very extended radio
continuum emission from Mrk 231 is given in Ulvestad et al. (1997b).
The VLA image from data taken in December 1996 shows a
peak surface brightness of 230 $\pm$ 8 mJy beam$^{-1}$ at 1.3$''$
resolution, while the data taken on August 31, 1997 show a peak surface
brightness of 242 $\pm$ 8 mJy beam$^{-1}$. 
Hence, the variation of the central continuum source over this time
period is less than 5$\%$ at 1.4 GHz. 

The VLBA image of Mrk 231 at 1.4 GHz at 6 mas resolution is shown in
Figure 2. The image shows a double source, with a compact central peak of 40
$\pm$ 4 mJy beam$^{-1}$, and a `lobe' extending 30 mas to the south with a
total flux density of 44 $\pm$ 5 mJy. There is also evidence for 
a fainter lobe extending about 20 mas to the north with a total flux
density of about 4 $\pm$ 1 mJy. This northern lobe has been detected
at a number of  frequencies in the sensitive VLBI continuum
observations of Ulvestad et al. (1997a).  The southern radio
lobe is thought to be the approaching lobe, based on the observation
of excess free-free absorption toward the northern lobe 
(Ulvestad et al.  1997a).  
The total flux density of the source in this image is 100 $\pm$ 10
mJy. There is a low surface brightness `bridge' connecting the 
central peak with the southern lobe, perhaps indicating
a jet.  A detailed discussion of the
continuum emission on 10 mas scales from Mrk 231 is given in Ulvestad
et al. (1997a).

The VLBA data also reveal the existence of diffuse emission on a scale
considerably larger than the compact core and lobes seen in Figure 2.
The shortest projected interferometer spacings of our VLBI experiment 
are about 180 k$\lambda$, where $\lambda$ is the observing wavelength,
corresponding to a resolution of about 1$''$, 
and the source flux density increases to about 230 mJy on these
spacings, comparable to the observed flux density on the longest VLA
spacings.  In order to determine the
distribution of this diffuse emission, we subtracted the CLEAN
component model of the source seen at high resolution in Figure 2 from
the visibility data set, and we  imaged the residual visibilities
using a Gaussian taper with FWHM = 3 M$\lambda$, corresponding to
data primarily from the inner-most antennas of the array
(the VLA, Pie Town, Los Alamos, and Fort Davis). The resulting image
is shown  in Figure 3 at a resolution of 60 mas. The image shows 
an elliptical source roughly centered on the 
core-lobe source, with a major axis of 440 mas oriented east-west,
and a minor axis of 310 mas. The total flux density in this diffuse
component is 130 $\pm$ 13 mJy. The average surface brightness 
in the inner 200 mas is 6$\pm$2 mJy beam$^{-1}$, corresponding  to a 
brightness temperature of 10$^6$ K.  The surface brightness then drops
to about 3 mJy beam$^{-1}$ at radii larger than 100 mas.
From the longest spacings of the VLA we derive an upper limit of
0.5$''$ to the size of this diffuse emission.

\subsection{HI 21cm Absorption: Systemic Velocity}

Figure 4 shows the HI 21cm absorption spectrum of Mrk 231 made
with the VLA  in August 1997 at a spectral resolution of 21 km s$^{-1}$ and
a spatial resolution of 18$''$. The continuum flux density 
of 242 $\pm$ 8 mJy has been subtracted, and the spectrum has been converted to
optical depth using this continuum level. The rms noise ($\sigma$) in this
spectrum is 0.8 mJy beam$^{-1}$ channel$^{-1}$. The peak 
absorbed flux density is  $-19\pm$1 mJy. A  single
Gaussian component fit to the absorption line results in a peak 
opacity of 0.083 $\pm$ 0.002, and a velocity FWHM = 179 $\pm$ 5 km
s$^{-1}$. The  peak of the Gaussian fit is at a heliocentric redshift of
0.04217 $\pm$ 0.000015, or cz = 12642 $\pm$ 4 km s$^{-1}$. We adopt this
redshift as the zero point in the velocity scale in this, and
subsequent, spectra. 

The most surprising results from this study are the HI 21cm
absorption spectra toward the VLBI core and lobe 
components seen on 10 mas scales in
Figure 2. These spectra are shown in Figure 5 at 56 km s$^{-1}$
resolution.  The spectra were extracted at the position of the peak 
surface brightnesses of the two continuum components on images 
convolved to 13 mas resolution. The spectra have been converted to
optical depth using the continuum surface brightnesses of 43 mJy
beam$^{-1}$ for the peak, and 20 mJy beam$^{-1}$ for the southern lobe.
The rms noise  in these spectra is 0.3 mJy beam$^{-1}$
channel$^{-1}$. We expected to 
see strong absorption (as much as  19 mJy), 
either toward the nucleus or toward the southern
lobe. However, no absorption is seen toward either of these components to
a 3$\sigma$ level of 1 mJy, or 
optical depth limits of 0.02 for the peak and 0.04 for the
southern lobe. The faintness of the northern lobe only allows us to
set a 3$\sigma$ optical depth limit of 0.5.

Spectral images were then synthesized by tapering the VLBI visibilities
to look for absorption against the diffuse  continuum emission seen on
scales of a few hundred  mas as shown
in Figure 3.  A spectrum made with a spatial resolution of 400 mas
(comparable to the size of the diffuse radio continuum 
emission shown in Figure 3), and
a spectral resolution of 28 km s$^{-1}$, is shown in Figure 6. 
The  peak surface brightness is 91 mJy beam$^{-1}$ (after subtracting the
contribution from the VLBI core and lobes). A strong HI 21cm absorption
line is observed, with a peak flux density of $-14\pm$2 mJy. A  single
Gaussian component fit to the absorption line results in a peak 
opacity of 0.17$\pm$0.02, a velocity FWHM = 193 $\pm$ 25 km
s$^{-1}$, and a velocity centroid of $-21\pm12$ km s$^{-1}$. 

Spectral channel images of the HI 21cm absorption toward the
diffuse continuum emission from the central regions of Mrk 231
are shown in Figure 7 for images tapered to a 
spatial resolution of 150 mas, and smoothed to a spectral resolution
of 58 km s$^{-1}$. The absorption shows a clear velocity gradient from
west to east, starting at $-$130 km sec$^{-1}$ at  100 mas
west of the continuum peak, and moving to +100 km sec$^{-1}$ at a
comparable distance to the east of the continuum peak. 

It is important to keep in mind that, unlike HI emission studies, 
HI absorption measurements depend both on the neutral gas
distribution and the background continuum  distribution. Hence it is
difficult to determine  whether the fading of the HI absorption signal
to the east and west is 
due to the lack of HI or the lack of continuum. 
We find that the peak opacity is roughly constant 
across the brighter regions of the source, with a value of about
0.18$\pm$0.04 to a distance of at 
least  100 mas both east and west of the nucleus. 
The continuum surface brightness beyond a radius of 100 mas is
too low to detect absorption in the current data.

\subsection{HI 21cm Absorption: The Broad Absorption Line Systems}

Optical spectra of Mrk 231 have revealed  broad
absorption line systems at heliocentric velocities of: cz = 4660 km
s$^{-1}$, 6250  km s$^{-1}$, and 8000 km s$^{-1}$ (Boroson et
al. 1991,  Forster et al. 1995, Rudy etal. 1985). The 4660 km
s$^{-1}$ and 6250  km s$^{-1}$ systems have velocity widths of about
200 km s$^{-1}$, while the system at 8000 km s$^{-1}$ has a velocity
width of about 1000 km  s$^{-1}$. These systems show absorption by low
ionization state species, such as NaI. Given that  NaI 
atoms have a lower ionization potential than HI, one might hope to
see associated absorption in the 21cm line of neutral hydrogen. 

Given the different velocity widths of the  optical BAL systems,
we observed the two lower redshift systems  with a velocity
resolution of 21 km s$^{-1}$, while the 
higher redshift system was observed with a velocity resolution of 87
km s$^{-1}$.  The absorption spectra  
are shown in Figure 8. No absorption was detected 
in any of the systems. The spectral rms for the two lower redshift
systems is 0.7 mJy channel$^{-1}$, while that in the spectrum of the
higher redshift system is 0.6  mJy channel$^{-1}$. The continuum flux
density is 240 $\pm$ 7 mJy, hence the 3$\sigma$ optical depth limits 
per spectral channel 
are 0.009 for the two lower redshift systems, and 0.008 for the
higher redshift system. 

\section{Discussion}

\subsection{The Neutral Atomic Disk}

The three important results from our high resolution imaging
observations  of the 
HI 21cm absorption at 12642 km s$^{-1}$  toward Mrk 231 
are that: (i)  absorption is {\sl not} 
seen toward the VLBI core and lobes,
(ii)  absorption  is detected toward a diffuse, elliptical radio
continuum component  which is centered on the 
nucleus and has a maximum size of 440 mas (= 260 h$^{-1}$ pc), and (iii) 
the absorption shows an east-west gradient in position and velocity
of about $\pm$ 110 km s$^{-1}$ out to radii $\approx$  100 mas (60
h$^{-1}$ pc). One  plausible explanation of these three
observed characteristics is that the neutral atomic gas, and the radio
continuum emitting gas, are in a disk 
centered on the active galactic nucleus (AGN), and
that the disk is oriented not too far from the sky plane, and that the
radio lobe axis is oriented perpendicular to the disk.
We discuss this model in the following section (see also Figure 9 and
the discussion at the end of section 4.2).

Perhaps the strongest argument that the absorbing neutral hydrogen
in Mrk 231 is in a disk is the simple fact that
such a gas disk has already been seen  in the CO molecular line images
of  Bryant and Scoville (1996), and that the parameters for HI
absorption are consistent with those of the molecular disk (see below).
The molecular observations show an
elliptical distribution for the CO, with a major axis of 1.2$''$
oriented east-west and a minor axis of 0.8$''$. The molecular gas
shows an east-west velocity gradient, ie. along the disk
major axis. The molecular disk model of
Bryant and Scoville is based primarily on the fact that, if the
molecular gas was spherically distributed about the nucleus, the gas
mass would greatly exceed the dynamical mass. They set 
a conservative upper limit to the disk inclination angle relative to
the sky plane of:~ $i$ $<$ 59$^o$. They also show that a
spherical distribution would  lead to an  A$_V$ $\approx$ 100 toward
the nucleus, in conflict with the observed value of A$_V$ $\approx$ 2
(Boksenberg et al. 1978). 

The HI observations presented herein show an 
east-west velocity gradient for the atomic gas that is similar in both
magnitude and position angle to  that seen for
the molecular disk, although on a factor three smaller
scale. We would then identify the observed HI gas with
the inner regions of the molecular  disk seen in CO emission.
We should emphasize that we cannot rule out an outflow 
model to explain the HI velocity gradient, as opposed to rotation. 
One argument against outflow is that the atomic and molecular gas
outflow axis would then be perpendicular (at least in projection) to
the radio jet outflow axis. 

An HI disk not too far from the sky plane  would provide
a natural explanation for the lack of HI absorption toward the VLBI
nucleus, and the southern (approaching) lobe.  Consider
the simple example of a constant opening angle disk in which the 
half width of the disk increases linearly with radial distance from
the center, with a constant of proportionality $C$. The nucleus will
be unobscured as long as:~ cotan($i$) $\le$ $C$. 
For instance, if we adopt a value of  $i$ = 45$^{o}$ (see section
4.2), then the nucleus and southern (approaching) lobe will be
unobscured as long as the disk half width is less than the disk
radius at any given radius. An additional factor
in forming the `HI hole' toward the nucleus of Mrk 231 could be
ionization of the inner disk by the active nucleus (section 4.2). 

The total column density for the HI 21cm absorption line
system seen toward the radio continuum disk in Mrk 231 is:~ N(HI) = 
6.3$\pm$0.7 $\times$10$^{19}$ $\times$ ${T_s}\over{f}$ cm$^{-2}$, 
where $T_s$ is the HI spin
temperature and $f$ is the HI covering factor (f $\le$ 1).
Assuming a pathlength of order 100 h$^{-1}$ pc through the
HI implies an average density of roughly:  n(HI) $\approx$ 0.3
$\times$ ${T_s}\over{f}$ h 
cm$^{-3}$.  The total atomic hydrogen mass in the disk is then of
order: M(HI) $\approx$  10$^{4}$  $\times$ ${T_s}\over{f}$ h$^{-2}$
M$_\odot$. The value of T$_s$ is unknown, but typical Galactic values
range from 10$^2$ to 10$^4$ K (Heiles and Kulkarni 1989). The value of
$f$ is also unknown, but if the radio continuum emitting regions are
co-spatial with the atomic gas (section 4.2), then  $f$ $\approx$ 0.5.

Bryant and Scoville (1996) estimate a minimum H$_2$
mass in Mrk 231 of about 10$^9$ M$_\odot$ within a radius of 300
h$^{-1}$ pc,  with an average  H$_2$ column density of 
10$^{23}$ cm$^{-2}$.  Roche and Chandler
(1993) show that the sub-mm and infrared spectrum of Mrk 231 can be
reasonably fit by a model of a dust disk with a diameter of
about  0.4$''$ and a temperature of 85 K. This disk would be
optically thick even in the far IR, with a
minimum total dust mass of 8x10$^{9}$ M$_\odot$.
Lastly, from the  free-free opacity toward the 
northern radio lobe in Mrk 231, Ulvestad etal. (1997a) estimate an
emission measure of 2$\times$10$^7$ pc cm$^{-6}$ toward the northern
lobe, implying a thermal electron column density of 10$^{23}$
cm$^{-2}$ on a scale of 20 mas.

If the  spin temperature of the neutral hydrogen is fairly high
($\ge$ 10$^3$ K), then the column densities in the ionized, neutral
atomic, and molecular  material are all of order 10$^{23}$ cm$^{-2}$,
with the different components being detected on increasing scales.
This leads to the interesting speculation that 
perhaps we are seeing the transition from an
ionized inner disk at radii $\le$ 20 h$^{-1}$ pc, to a neutral atomic
disk at radii between 20 h$^{-1}$ pc and 100 h$^{-1}$ pc, to a mostly
molecular and dusty disk at radii between 100 h$^{-1}$ pc and 300
h$^{-1}$ pc (Figure 9). Using the rough value for the mean n(HI) given
above, and assuming a neutral gas temperature of 10$^4$ K, the implied
Stromgren radius is: 
$r_{st}$ = 25 $\times$ [${L_{uv}}\over{10^{44}}$]$^{1/3}$
h$^{-2/3}$ pc, where $L_{uv}$ is the AGN luminosity  above 13.6 eV
in ergs s$^{-1}$. 

The dynamical mass can be estimated assuming that the HI is in a 
Keplerian rotating disk.
The observed velocity at a radius of 50 h$^{-1}$ pc is about 110 km
s$^{-1}$. Hence, the enclosed mass inside this radius is:
M$_{dyn}$ = 1.4$\times$10$^{8}$ h$^{-1}$ (sin $i$)$^{-2}$ M$_\odot$. 
Assuming that  $i$ $\approx$ 45$^o$ (see section 4.2),
leads to M$_{dyn}$ = 3$\times$10$^{8}$ h$^{-1}$
M$_\odot$.  Note that the HI absorption line FWHM 
of 180 km s$^{-1}$ seen by the shorter spacings of our  VLBI
observations  is  comparable to the value seen for the integrated 
CO emission from Mrk 231. Given the different maximum radii sampled in
these two observations (120 h$^{-1}$ pc versus 360 h$^{-1}$ pc), 
this similarity in line widths suggests that the rotation curve 
of the disk could be flat, or falling, at radii larger than about 100
h$^{-1}$ pc. In this case, we can set an upper limit to the velocity
dispersion in the disk of $\approx$ 140 km s$^{-1}$, and using the
equations of hydrostatic equilibrium (Scoville et al. 1997), we can
set an upper limit to the effective thickness of the disk of 
30 pc. For comparison, Scoville et al. (1997) derive a velocity
dispersion for the disk in Arp 220 of 90 km s$^{-1}$, from which they 
derive a disk thickness of 16 kpc.

\subsection{The Synchrotron Emitting Disk}

The diffuse radio continuum component seen by the shorter
VLBA spacings  (Figure 3) is elliptical with 
a major axis orientation  the same as that seen for 
the molecular gas (east-west), although on a factor three
smaller scale. The major-to-minor axis ratio for this
diffuse radio component is about 1.4, which is 
also similar to that of the CO disk. 
It is plausible, therefore, to assume that this radio continuum
emission is associated with the disk. 
For a circular, flat disk, the observed axis ratio implies a disk
orientation angle relative to the sky plane of 45$^o$.

The radio continuum `disk' in Mrk 231 has a brightness temperature of
10$^6$ K, suggesting  non-thermal emission, ie. synchrotron radiation.
One possible origin  for the relativistic electrons is first order 
Fermi acceleration in supernova remnant shocks due to massive star
formation in the disk. The proposed physical conditions in the disk
are certainly conducive to star formation
(Bryant and Scoville 1996, Scoville et al. 1997). A
diagnostic as to whether star 
formation is a reasonable explanation for the radio continuum emission
from the disk is the IR-to-radio flux density ratio:
Q $\equiv$ log [${S_{100} + 2.6 \times S_{60}}\over{3 \times
S_{1.4}}$] (Condon 1992), where $S_{1.4}$ is the radio continuum flux
density at 1.4 GHz in Jy, and $S_{60}$ and $S_{100}$ are the IR flux
densities at 60$\mu$m and 100 $\mu$m, respectively, also in Jy. 
The majority of the IR emission from Mrk 231 comes from a region with
a scale comparable to that of the radio continuum disk (Roche and
Chandler 1993, Mathews et al. 1987), with flux densities of: $S_{60}$
= 35 Jy, and $S_{100}$ = 32 Jy (Soifer et 
al. 1989). The radio continuum emission from the disk has:
$S_{1.4}$ = 0.13 Jy, leading to a value of:
Q = 2.5. This value is consistent with the tight
correlation  seen for starburst galaxies, for  which 
Q = 2.3 $\pm$ 0.2 (Condon 1992). The implied massive star formation
rate is 60 M$_\odot$ year$^{-1}$ in the disk for stars with mass $\ge$
5 M$_\odot$ using the empirical relations given in Condon (1992).  


The minimum pressure in the relativistic particles and
magnetic fields in the radio continuum disk of  
Mrk 231 is about  3$\times$10$^{-9}$ dynes cm$^{-2}$, 
with corresponding magnetic fields of order 300 $\mu$G. While this pressure
is much higher than typical interstellar pressures in our galaxy, 
it is comparable to pressures hypothesized in the nuclei of some
nearby starburst galaxies (Carilli 1996, Heckman
etal. 1990). Also,  this pressure is still a factor of a few below
that hypothesized by Bryant and Scoville (1996) in the
self-gravitating disk in Mrk 231. The synchrotron and inverse Compton
radiative  lifetimes for the relativistic
particles emitting at 1.4 GHz in the  disk of Mrk 231 are $<$ 
10$^5$ years.  Streaming velocities for relativistic electrons
in the interstellar medium are thought to be limited to the Alfven
speed due to scattering off self-induced Alfven waves (Wentzel 1974). 
Assuming a mean thermal particle density of order 10$^3$ cm$^{-3}$
implies an Alfven speed of order 20 km s$^{-1}$, leading to a
streaming distance less than a few pc in the particle lifetime. Hence
{\sl in situ} particle acceleration may be required, such as could
occur eg. in supernova remnant shocks (Condon et al. 1991). 

The physical conditions for the thermal and non-thermal gas in
the sub-kpc disk of Mrk 231 are similar to those proposed for
compact nulcear starburst galaxies by Condon et al. (1991). In a sample of
39 IR luminous galaxies, Condon et al. find that about a third show
evidence for a nuclear  starburst region with star formation rates $\ge$ 10
M$_\odot$ year$^{-1}$ on scales of only a few
hundred parsecs. They propose that the sources are optically thick in the
IR, and that the free-free opacities can be significant at 
cm wavelengths. The expected supernova rates are about one per year, 
leading to very high interstellar pressures in theses
regions ($\approx$ 10$^{8}$ dyne cm$^{-2}$). A potential problem with
interpreting the disk in Mrk 231 as a compact starburst region of the
type seen by Condon et al. is the high radio continuum brightness
temperature of 10$^6$ K at 1.4 GHz. 
Condon et al. use an empirical  scaling relation between thermal and
non-thermal radio continuum emission based on more extended starburst
galaxies to set an upper limit to the allowed brightness temperature
in a compact starburst of 10$^5$ K at 1.4 GHz assuming a thermal
electron temperature of 10$^4$ K. A possible solution to this
problem is to hypothesize that the emipirical scaling relations break
down under the extreme conditions in compact starbursts. An argument
against this solution is the fact that these regions obey the 
standard radio-IR correlation  for starburst galaxies. 
Alternatively, one could arbitrarily raise the thermal electron
temperature by an order of magnitude, or else assume that the 
relativistic electrons are not accelerated in supernova remnant shocks.
A possible alternative particle acceleration mechanism is 
stochastic  (second order) Fermi acceleration in a turbulent disk.

Recent observations have supported the hypothesis of sub-kpc gas disks in
the nuclei of two other galaxies.
Scoville et al. (1997) have used high resolution CO observations to
infer a molecular disk in the ULIRG Arp 220 with a mass,  size, and
velocity structure similar to that seen in Mrk 231. They suggest that
the disk is fairly thin, with a diameter of 220 pc and a scale
height of 16 pc, based on the observed line widths and 
assuming hydrostatic equilibrium. While they find it
difficult to model the IR emission from  Arp 220 via star formation
alone, they also point out that, given the physical conditions in the
disk, it would be hard to avoid active star formation, perhaps at a 
rate as high as 90 M$_\odot$ year$^{-1}$. Mundell et al. (1995) 
have proposed a gaseous disk in the nucleus 
of the Seyfert II galaxy NGC 4151, although on a factor of a few
smaller scale than the disk in Mrk 231, based on high resolution HI
21cm observations (see also Ulvestad et al. 1997c, and Silver, Taylor, and
Vermeulen 1997). 

Figure 9 shows a schematic diagram of the model presented above to
explain the various observations of the nuclear regions of Mrk 231. 
We hypothesize a gaseous disk comprised of multiple phases:  
(i) ionized atomic gas, as seen in
free-free absorption on scales $\approx$ 20 h$^{-1}$ pc, 
(ii) neutral atomic gas, as seen in HI 21cm
absorption on scales $\approx$ 120 h$^{-1}$ pc, (iii) relativistic
gas, as seen in radio continuum emission on scales $\approx$ 240 pc, 
and (iv) molecular gas and dust, as seen in CO and infrared emission
on scales $\approx$ 600 h$^{-1}$ pc.\footnote{Note that the scales quoted
above are set mainly by observational limitations, and hence 
the data do not preclude a mixture  of the
various gaseous phases on  different scales.}
This disk is centered on the AGN, and is oriented at about
45$^o$ from the sky plane, and the radio jet axis is oriented
perpendicular to the disk.
The southern (brighter) pc-scale radio lobe
is approaching, and hence is unobscured by the disk.
In order to have an unobscured view into
the nucleus (in terms of HI 21cm absorption), the inner disk opening
half-angle must be less than the angle of the disk relative to our
light sight (45$^o$), although ionization of the inner disk 
by the AGN could weaken this constraint. In this model the neutral atomic
gas is roughly co-spatial  with the diffuse radio continuum disk,
thereby giving rise to the 
HI 21cm absorption seen by  the VLA and the shorter VLBI baselines, 
while at the same time 
the radio nucleus and the southern pc-scale lobe are not obscured by the
HI disk. The diffuse radio continuum emission from the disk is likely
to be driven by distributed  massive star formation. An upper limit
to the velocity dispersion in the disk provides an upper limit to the
disk effective thickness of 30 pc.

\subsection{The Temperature of the BAL Gas}

No HI 21cm absorption has been detected in the velocity
range  of the optical low ionization BAL systems. The strongest and
broadest of the BAL systems toward Mrk 231 is the one at cz = 8000 km
s$^{-1}$. This optical system has been analyzed extensively by Forster
etal. (1995), who find that the broad NaI line can be reasonably fit by 
a blend of nine unsaturated absorbing clouds with a total velocity spread
of 840 km s$^{-1}$,  with optical depths varying between 0.1 and 1.5,
and Doppler parameters between 50 km s$^{-1}$ and 70 km s$^{-1}$. The 
implied NaI column density in each component is about: N(NaI) =
2$\times$10$^{13}$  cm$^{-2}$, with the strongest component having a
column density of 6$\times$10$^{13}$ cm$^{-2}$.  

Our 3$\sigma$ neutral hydrogen column density limit towards this system 
is: N(HI) $\le$ 1.3$\times$10$^{18}$ ${T_s}\over{f}$ cm$^{-2}$. 
If the BAL gas has a solar 
abundance of NaI, then the implied HI column density would be: N(HI) = 
3.6$\times$10$^{19}$ $\times$ $d$ cm$^{-2}$, where $d$ is the dust
depletion factor ($d$ $\ge$ 1). The implied lower limit to the spin
temperature is then: $T_s$ $\ge$ 30 $\times ~d~\times~f$ K.
Boksenberg etal. have hypothesized that the two magnitudes of visual
extinction   toward the nucleus of  Mrk 231 is due to the broad 
absorption line systems. If this is the case, then one might expect~
$d$ $\approx$ 10 (Spitzer 1978). Alternatively, 
assuming a Galactic dust-to-gas ratio, the observed
value of A$_v$ = 2 implies a total N(HI) = 3.2$\times$10$^{21}$
cm$^{-2}$. Dividing this equally among the nine velocity components, and
comparing the result to the HI 21cm absorption measurements leads to a
lower limit to the HI spin temperature of:  $T_s$ $\ge$ 275 $\times~ f$ K. 
Note that the BAL gas must cover at least the active nucleus, which
sets a lower limit: $f$ $>$ 0.2,
based on the VLBI imaging results, leading to $T_s$ $>$ 50 K.

Given the detection of low ionization species such as NaI and CaII in
the BAL systems toward the nucleus of Mrk 231, we had hoped to detect
HI 21cm  absorption. The negative results in the radio  require that
the neutral atomic gas be fairly warm,
unless the  NaI abundance is higher than
solar, or the dust-to-gas ratio is higher than Galactic, or
the observed extinction toward the nucleus of Mrk 231 is not
due to the BAL gas.

\section{Summary}

Our imaging with the VLBA of the neutral
hydrogen 21cm absorption line system seen toward the nuclear regions
of Mrk 231 has shown that the absorption is not toward the 
radio core or lobes seen on pc-scales, but that the absorption occurs
against a more diffuse radio continuum component seen on a scale of
a few hundred parsecs. A plausible explanation for these
data is that both the HI 21cm absorption  and the radio
continuum emission occur in a disk centered on the active galactic
nucleus. The disk cannot be oriented too far from the sky plane,
or else the  nucleus would show HI 21cm absorption, and the
pc-scale radio lobe source must be oriented close to the axis of the disk.
The radio continuum morphology is consistent with a disk of
maximum radius of 440 mas  (260 h$^{-1}$ pc) at an
inclination angle of 45$^o$, with a major axis oriented east-west.
The HI 21cm absorption shows an east-west gradient in position and velocity
of about $\pm$ 110 km s$^{-1}$ out to radii $\approx$  100 mas (60
h$^{-1}$ pc). The parameters for the morphology of this
disk (orientation of the major axis and inclination angle),
and the observed HI velocity field, are very similar to those seen
for the self-gravitating 
molecular gas disk discovered  by Bryant and Scoville (1996) in 
Mrk 231, although on a factor three smaller
scale. We would then identify the observed neutral atomic and radio
continuum emitting disk  with the inner regions of the 
molecular  disk seen in CO emission. 

The physical conditions for the thermal and
non-thermal gas in the sub-kpc disk of Mrk 231 are similar to those
proposed for compact nuclear starburst galaxies (Condon et al. 1991),
and in particular to the conditions proposed for the sub-kpc gas disk
in Arp 220 (Scoville, Yun, and Bryant 1997). The minimun pressure in the 
non-thermal gas is 3$\times$10$^{-9}$ dynes cm$^{-2}$, 
with corresponding magnetic fields of order 300 $\mu$G.
The  IR-to-radio flux density ratio for the disk implies a value of
Q = 2.5, consistent with the tight correlation  seen for starburst
galaxies. The implied massive star formation  rate in the disk
is 60 M$_\odot$ year$^{-1}$. From the neutral hydrogen velocity field we
derive a gravitational mass enclosed within a 50 
h$^{-1}$ pc radius of 3$\times$10$^{8}$ h$^{-1}$ M$_\odot$. 
We propose that Mrk 231 provides a relatively nearby example of the
physical processes and conditions seen in high redshift ULIRGs
(Scoville, Yun, and Bryant 1996).

We also present a search for HI 21cm absorption associated with the
optical broad absorption line (BAL) systems seen toward Mrk 231. 
We do not detect HI 21cm absorption associated with any of
the optical BAL systems. The 3$\sigma$ optical depth limit
is 0.009 for the systems at cz = 4660 km s$^{-1}$ and
6250 km s$^{-1}$, with a velocity resolution of 21 km s$^{-1}$.
The limit for the broadest system at  8000 km s$^{-1}$ is 0.008 at a
velocity resolution of 84 km s$^{-1}$. 
These negative results require  that 
the neutral atomic gas in the BAL clouds  be fairly warm
(T$_s$ $>$ 50 K), unless the  NaI abundance is higher than
solar, or the dust-to-gas ratio is higher than Galactic, or
the observed extinction toward the nucleus of Mrk 231 is not
due to the BAL gas.

\vskip 0.2truein

We thank Min Su Yun, Alan Roy, and Greg Taylor for useful comments and
discussions. 
This research made use of the NASA/IPAC Extragalactic Data Base (NED)
which is operated by the Jet propulsion Lab, Caltech, under contract
with NASA. The National Radio Astronomy Observatory is a facility of
the National Science Foundation operated under cooperative 
agreement by Associated Universities, Inc. 

\newpage

\centerline{\bf References}

Armus, L., Surace, J.A., Soifer, B.T., Matthews, K., Graham, J.R., and
Larkin, J.E. 1994, A.J., 108, 76

Bocksenberg, A., Carswell, R.F., Allen, D.A., Fosbury, R.A., Penston,
M.V., and Sargent, W.L.W. 1977, M.N.R.A.S. 178, 451

Boroson, T.A., Meyers, K.A., Morris, S.L., and Persson, S.E. 1991,
Ap.J. (letters), 370, L19

Bryant, P.M. and Scoville, N.Z. 1996, Ap.J., 457, 678

Carilli, C.L., Perley, R.A., Dreher, J.W., and Leahy, J.P. 1991,
Ap.J., 383, 554

Carilli, C.L. 1996, A.\&A., 305, 402

Clark, B.G. 1980, A\&A, 89, 377

Condon, J.J. 1992, A.R.A.A., 30, 575

Condon, J.J., Huang, Z.P., Yin, Q.F., and Thuan, T.X. 1991, Ap.J. 378, 65



Dickey, J. 1982, Ap.J., 263, 87

Forster, Karl, Michael, Rich R., and McCarthy, J.K. 
1995, Ap.J., 450, 74


Hamilton, J.B. and Keel, W.C. 1987, Ap.J.,  321, 211

Heckman, T.M., Balick, B., and Sullivan, W.T. 1978, Ap.J. 224, 745

Heckman, T.M., Armus, L., and Miley, G. 1990, Ap.J. (supplement), 73,
883 

Heiles, C. and Kulkarni, S. 1989, in {\sl Galactic and Extragalactic Radio
Astronomy}, eds. K. Kellerman and G. Verschuur (Berlin: Kluwer)

Hutchings, J.B. and Neff, S.G 1987, A.J., 93, 14

Kollatschny, W., Dietrich, M., Hagen, H. 1992, A\&A (letters), 264, L5

Matthews, K., Neugebauer, G., McGill, J. and Soifer, B.T. 1987,
A.J. 94, 297

Mundell, C.G., Pedlar, A., Baum, S.A., O'Dea, C.P., Gallimore, J.F.,
and Brinks, E. 1995,  M.N.R.A.S., 272 355

Napier, P.J., Thompson, A.R., and Ekers, R.D. 1983, Proc. IEEE, 71,
1295 

Napier, P.J., Bagri, D.S., Clark, B.G., Rogers, A.E.E., Romney, J.D., 
Thompson, A.R., and Walker, R.C. 1994, Proc. IEEE, 82, 658

Roche, P.F. and Chandler, C.J.  1993, M.N.R.A.S., 265, 486

Rudy, R.J., Foltz, C.B, and Stocke, J.T. 1985, Ap.J., 288, 531

Sanders, D.B. and Mirabel, I.F. 1996, A.R.A.A., 34, 749

Sanders, D.B., Soifer, B.T., Elias, J.H., Madore, B.F., Matthews, K.,
Neugebauer, G., and Scoville, N.Z. 1988, Ap.J., 325, 74

Silver, C.S., Taylor, G.B., and Vermeulen, R.C. 1997, Ap.J. (letters),
submitted 

Soifer, B.T., Boehmer, L., Neugebauer, G., and Sanders, D.B. 1989,
A.J., 98, 766

Scoville, N.Z., Yun, M.S., and Bryant, P.M. 1997, Ap.J., 484, 702

Scoville, N.Z., Yun, M.S., and Bryant, P.M. 1996, in {\sl Cold  Gas at
High Redshift}, eds. M. Bremer, P. van der Werf, H. R\"ottgering, and 
C. Carilli (Kluwer: Dordrecht), p. 25

Spitzer, Lyman 1978, {\sl Physical Processes in the Interstellar
Medium}, (New York: Wiley)

Surace, J.A., Sanders, D.B., Vacca, W.D., Veilleux, S., and
Mazzarella, J.M. 1997, Ap.J., in press

Ulvestad, J.S., Wrobel, J.M., and  Carilli, C.L. 1997a, in {\sl IAU
Colloquium 164: Radio Emission from Galactic and   Extragalactic
Compact Sources}, eds. G.B. Taylor, J.M. Wrobel  \& J.A. Zensus (ASP: San
Franscico)

Ulvestad, J.S., Wrobel, J.M., and  Carilli, C.L. 1997b, in preparation

Ulvestad, J.S., Roy, A.L., Colbert, E.J.M., and Wilson, A.S. 1997c,
Ap.J., submitted


Walker, R.C. 1985, in {Aperture Synthesis in Radio Astronomy}, eds. 
R. Perley, F. Schwab, and A. Bridle (NRAO: Green Bank), p. 189

Weedman, D.W. 1983, Ap.J., 243, 756 

Wentzel, D.G. 1974, A.R.A.A., 12, 71

\newpage

\centerline{Figure Captions}

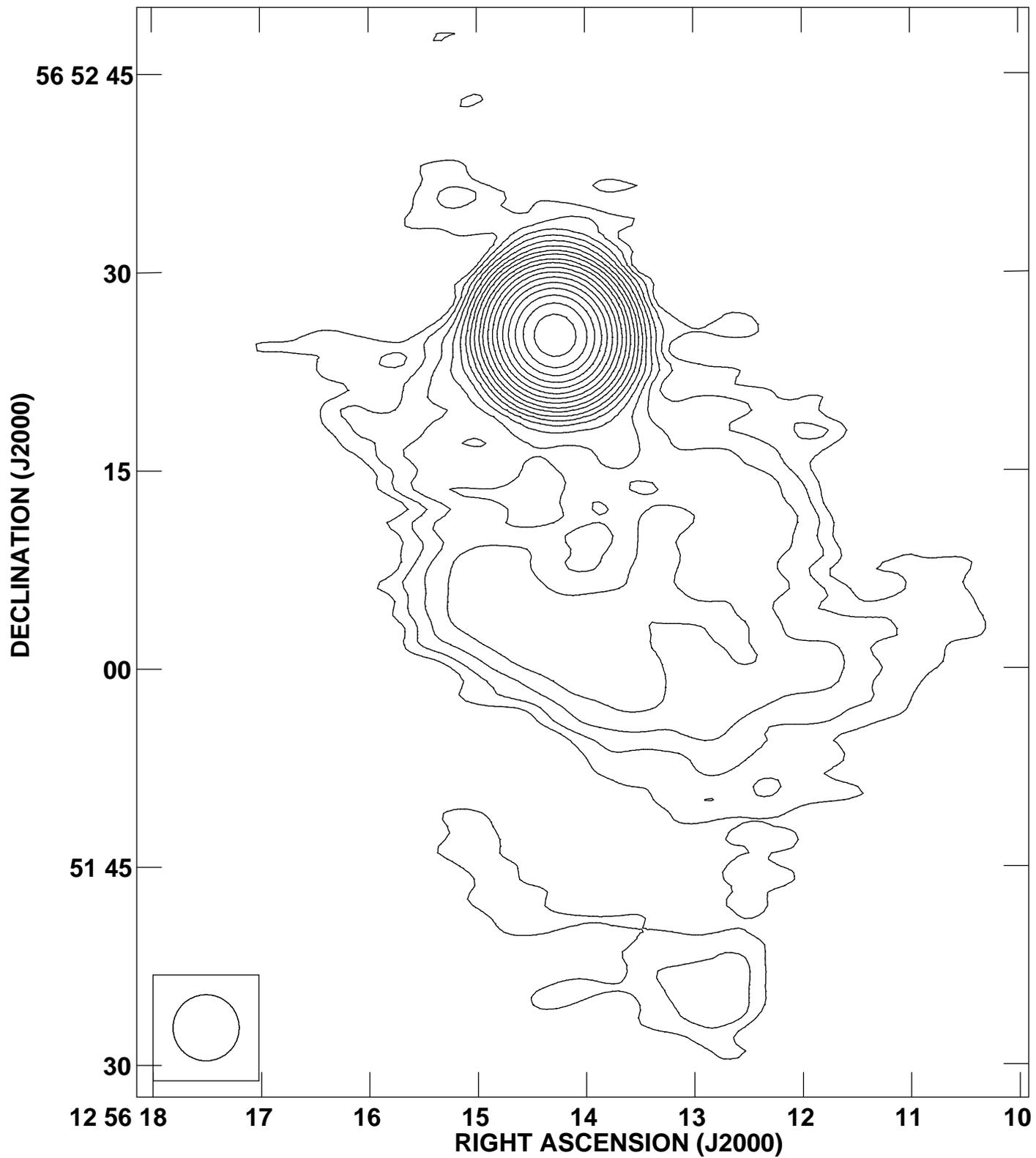
\noindent Figure 1 -- An image of Mrk 231 at 1.4 GHz with a resolution
of 5$''$ made with the VLA from data taken on January 27, 1993.
The contour levels are a geometric progression in the square root of
two, hence every two contours implies a factor two rise in 
surface brightness. The first contour level is 0.5 mJy beam$^{-1}$,
and the peak surface brightness is 240  mJy beam$^{-1}$. 

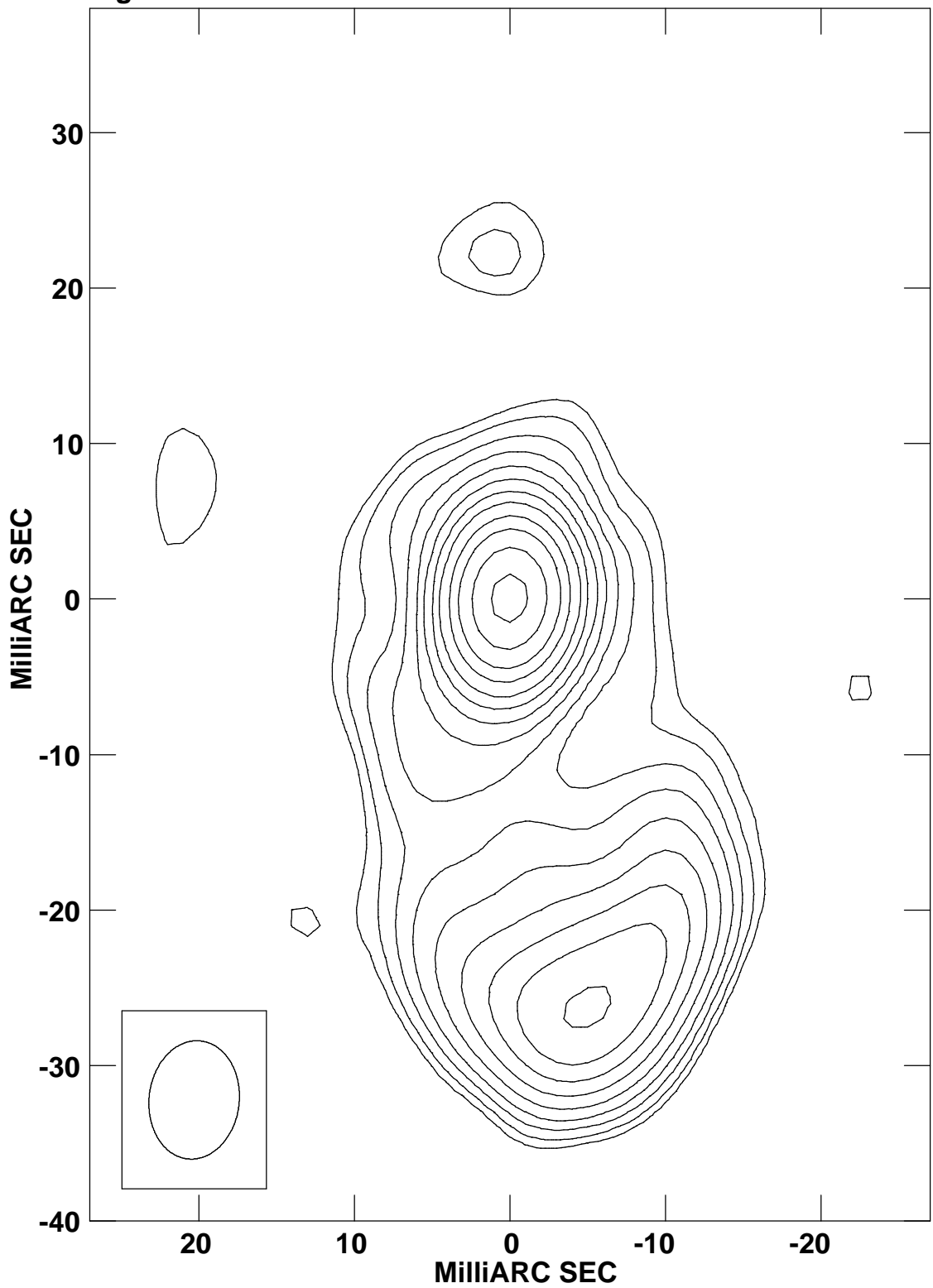
\noindent Figure 2 -- An image of Mrk 231 at 1.4 GHz with a resolution
of 8$\times$6 mas  made with the VLBA from data taken on December 27, 1996.
The contour scheme is the same a Figure 1, with the 
first contour level equal to 0.8 mJy beam$^{-1}$,
and the peak surface brightness equal to 40  mJy beam$^{-1}$. 

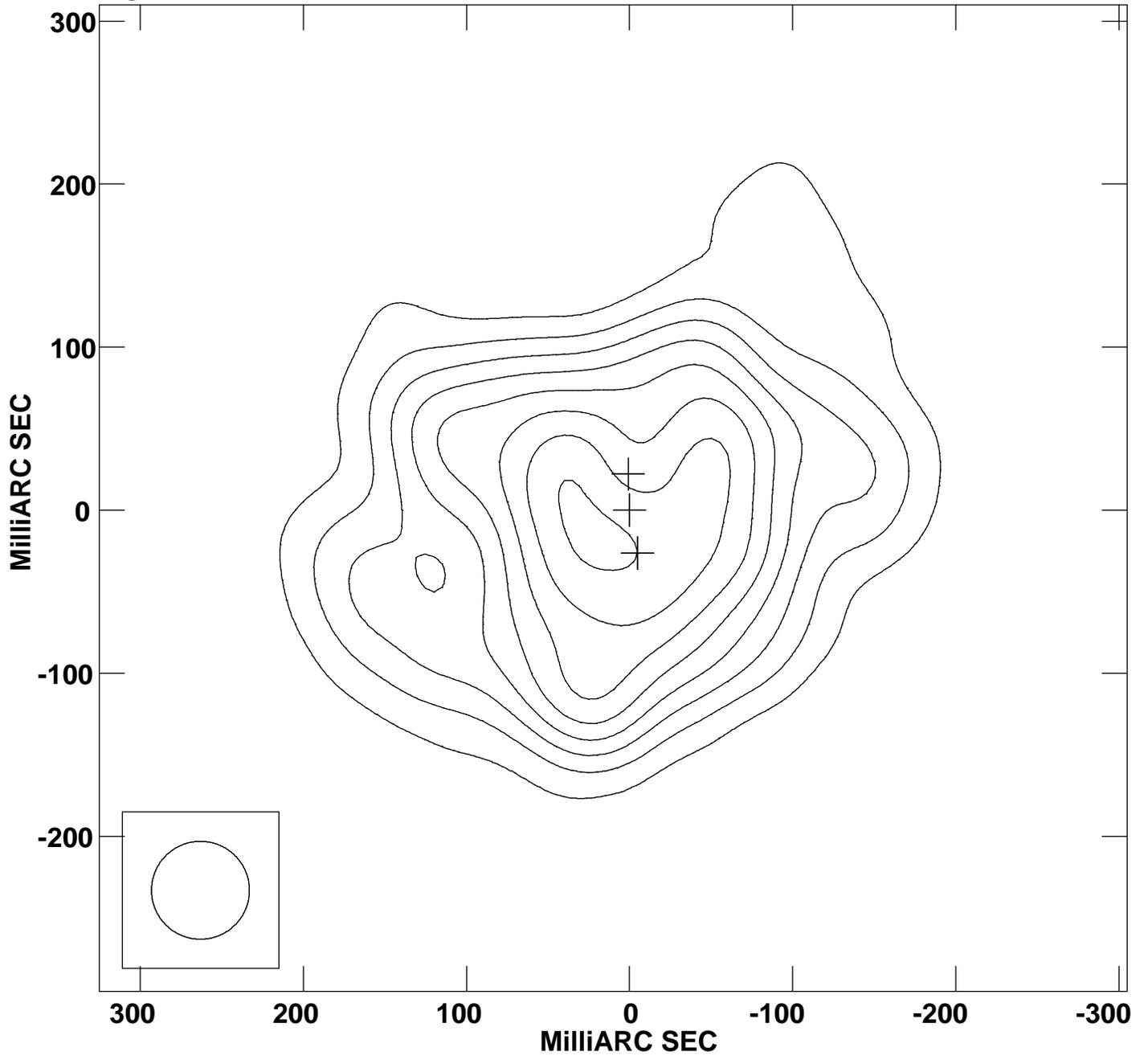
\noindent Figure 3 -- An image of the radio continuum emission 
from Mrk 231 at 1.4 GHz with a resolution
of 60 mas made with the VLBA from data taken on December 27, 1996.
A CLEAN component model of the nucleus and pc-scale lobes (100 mJy
total) as seen in Figure 2 has been subtracted from the visibilities 
before imaging. The contour 
levels are:~ 1, 2, 3, 4, 5, 6, 7, 8 mJy beam$^{-1}$,
and the peak surface brightness is 8.2 mJy beam$^{-1}$. 
The crosses mark the positions of the  nucleus, southern lobe, and
northern lobe as seen in Figure 2. 

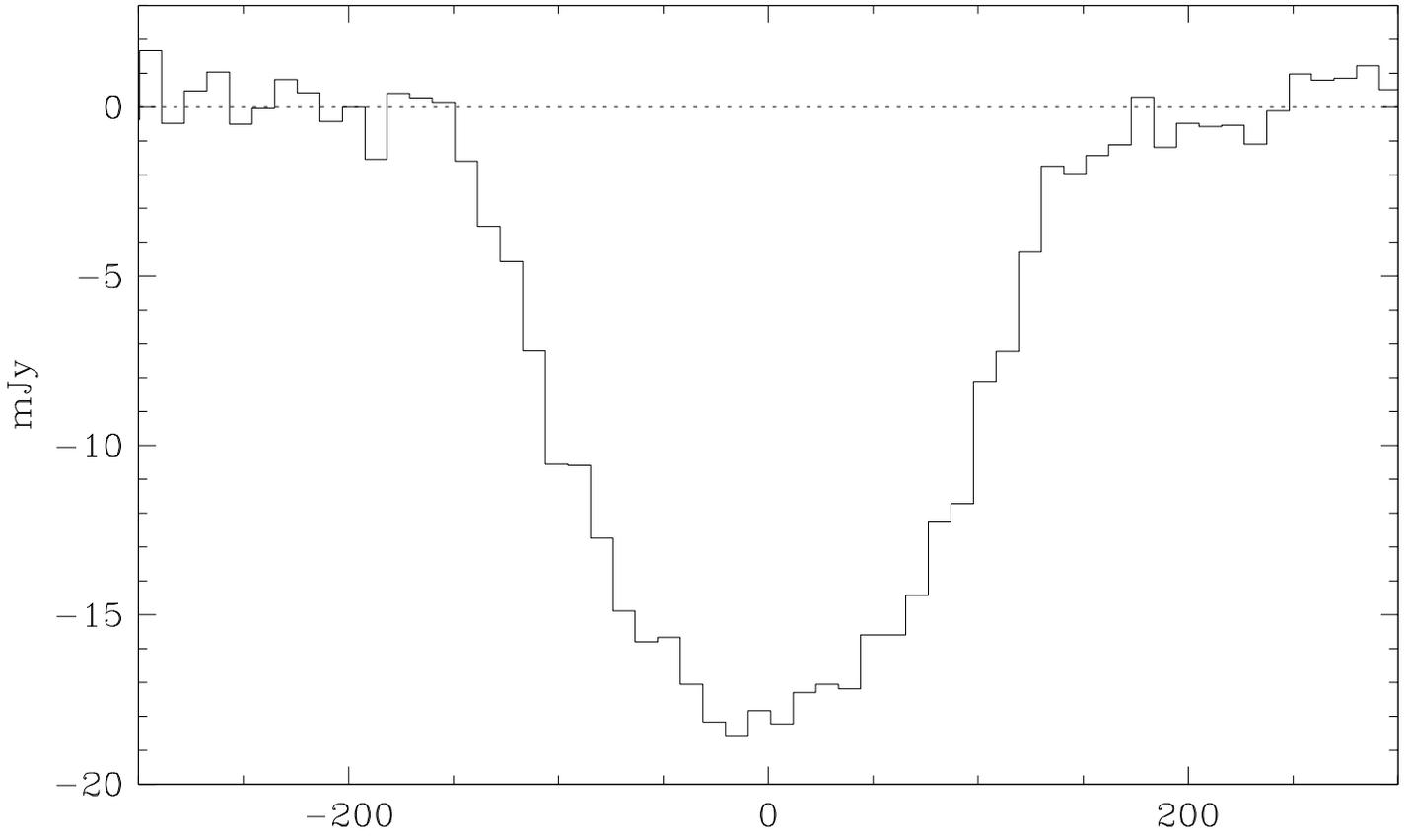
\noindent Figure 4 -- The HI 21cm absorption spectrum of Mrk 231
made with the VLA on August 31, 1997. The zero point on the velocity
scale corresponds to a heliocentric redshift of 
0.04217 $\pm$ 0.000015, or cz = 12642 $\pm$ 4 km s$^{-1}$. 
The spatial resolution is 18$''$ and the spectral resolution is 10.7
km s$^{-1}$ channel$^{-1}$. The upper plot shows the absorbed flux
density, after subtracting the continuum flux density of 242 mJy, and
fitting a linear baseline to off-line channels. The lower plot shows
the same spectrum converted to optical depth using the continuum flux
density. The dashed-line is a one component Gaussian model fit 
with a peak opacity of 0.083 $\pm$ 0.002, 
and a velocity FWHM = 179 $\pm$ 5 km s$^{-1}$.

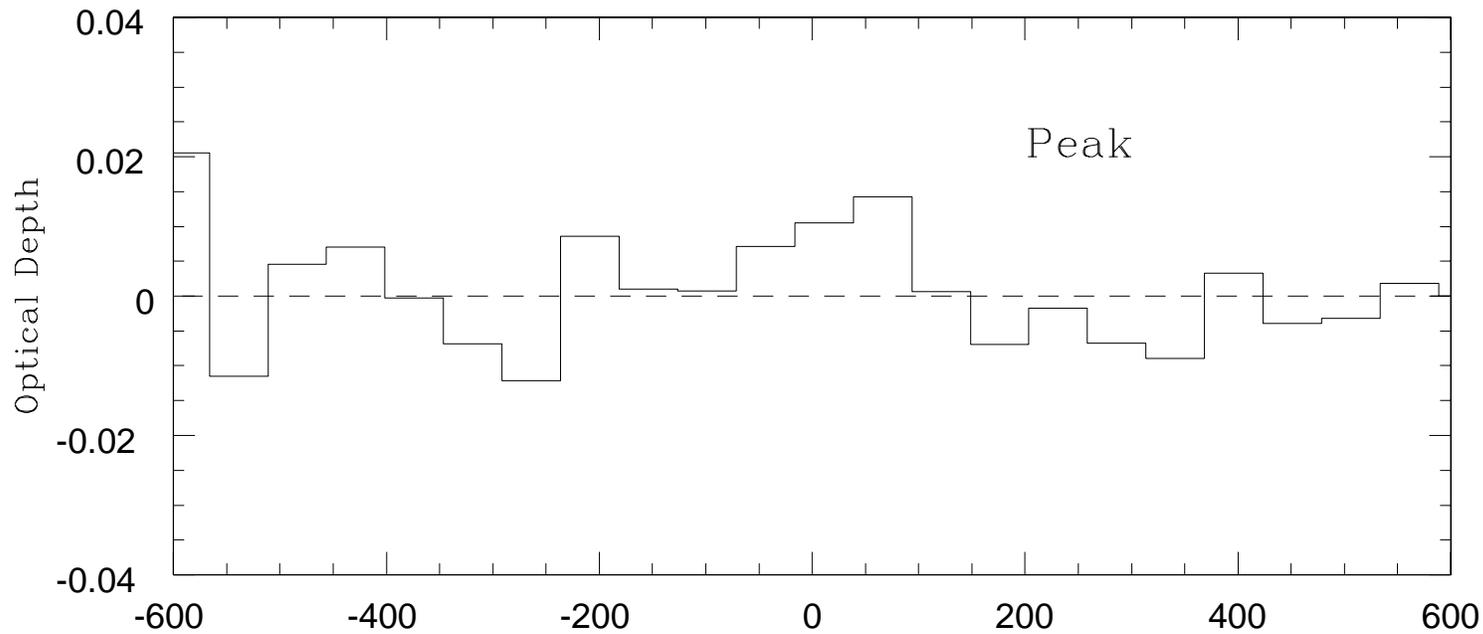
\noindent Figure 5 -- The HI 21cm absorption spectra toward the
pc-scale VLBI components  in Mrk 231
made with the VLBA on December 27, 1996. The zero point on the velocity
scale corresponds to a heliocentric redshift of 
0.04217 $\pm$ 0.000015, or cz = 12642 $\pm$ 4 km s$^{-1}$. 
The spatial resolution is 13 mas and the spectral resolution is 56
km s$^{-1}$ channel$^{-1}$. The upper plot shows the optical depth 
spectrum toward the peak surface brightness of the nucleus
of 43 mJy beam$^{-1}$. The middle plot shows the optical depth 
spectrum toward the southern lobe with a surface brightness of 20 mJy
beam$^{-1}$. 

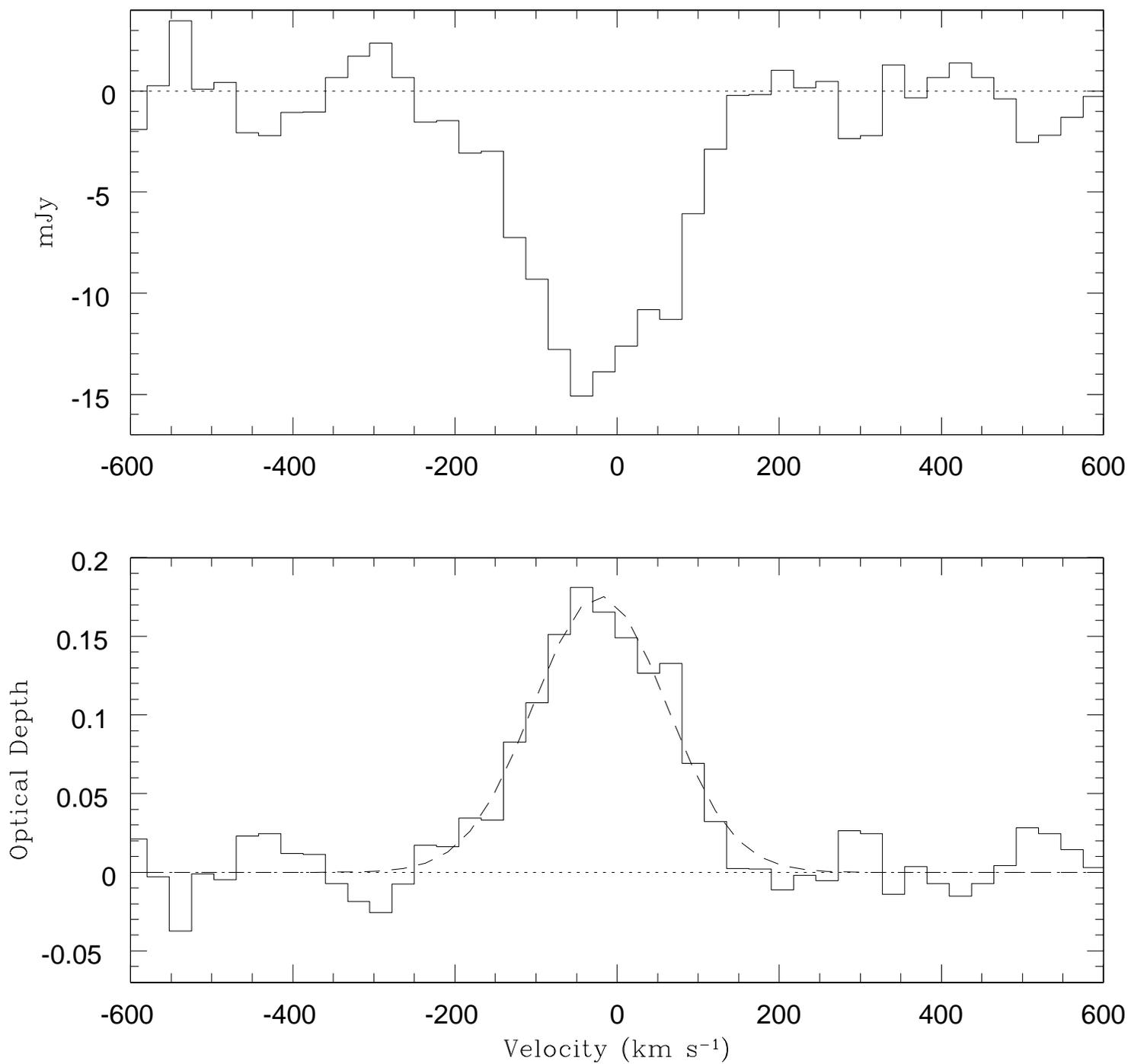
\noindent Figure 6 -- The HI 21cm absorption spectrum toward
radio continuum `disk' in Mrk 231 (see Figure 3)
made with the VLBA on December 27, 1996. The zero point on the velocity
scale corresponds to a heliocentric redshift of 
0.04217 $\pm$ 0.000015, or cz = 12642 $\pm$ 4 km s$^{-1}$. 
The spatial resolution is 0.4$''$ and the spectral resolution is 28
km s$^{-1}$ channel$^{-1}$. The upper plot shows the absorbed flux
density, after subtracting the continuum flux density of 91 mJy, and
fitting a linear baseline to off-line channels. The lower plot shows
the same spectrum converted to optical depth using the continuum flux
density. The dashed-line is a one component Gaussian model fit 
with a peak opacity of 0.17$\pm$0.02, a velocity FWHM = 193 $\pm$ 25 km
s$^{-1}$, and a velocity centroid of $-21$ $\pm$ 12 km s$^{-1}$. 

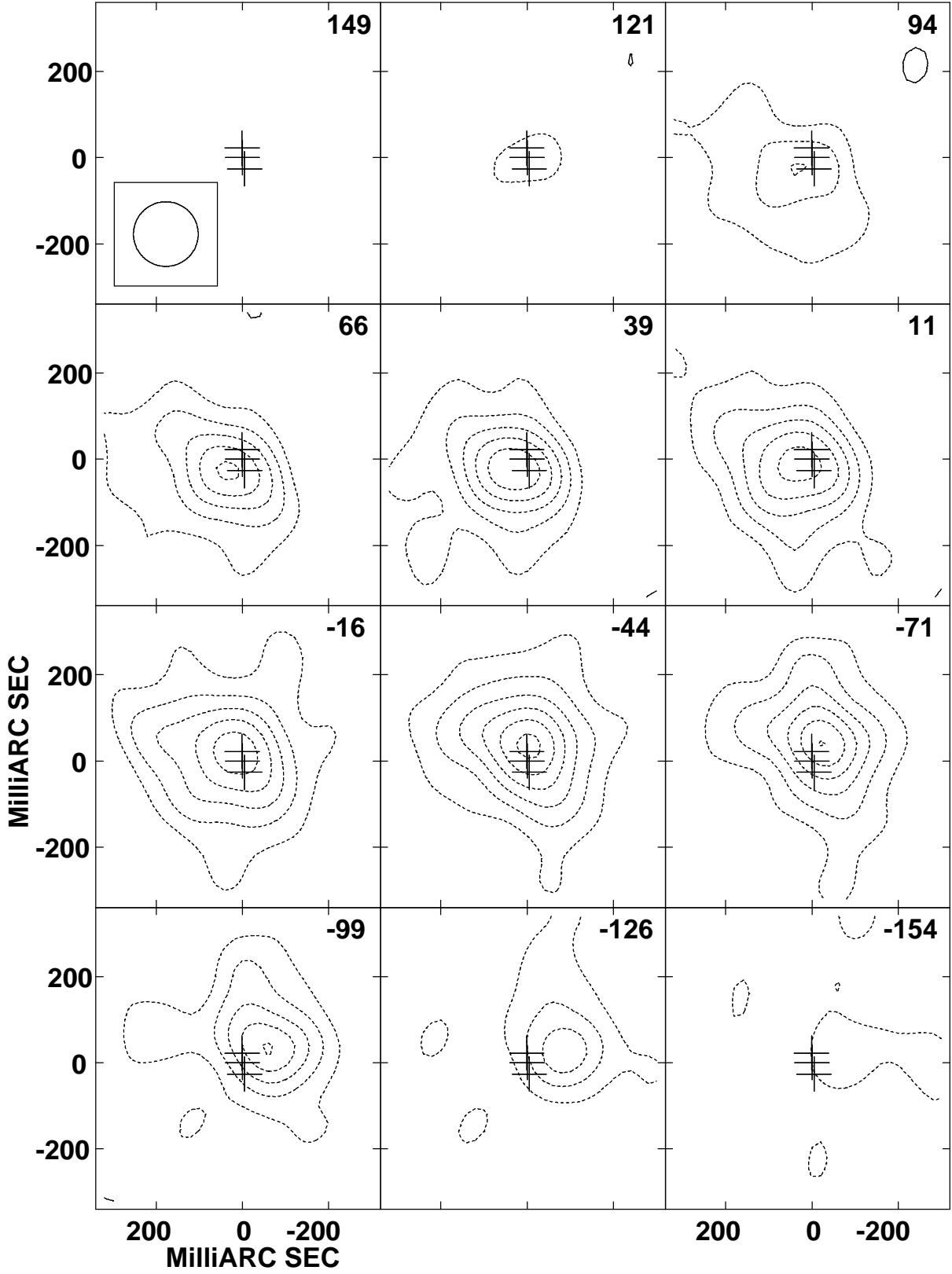
\noindent Figure 7 -- Spectral channel images of the HI 21cm
absorption toward Mrk 231 made from data taken with the VLBA on
December 27, 1996.  The channel spacing is 28 km s$^{-1}$, however the
data have been Hanning smoothed to 56  km s$^{-1}$ channel$^{-1}$,
hence adjacent channels are not independent. The spatial resolution is
150 mas. Each panel is labeled with the velocity relative to
heliocentric redshift of  0.04217 $\pm$ 0.000015, or cz = 12642 $\pm$
4 km s$^{-1}$.  The crosses mark the positions of the  nucleus,
southern lobe, and northern lobe as seen in Figure 2. The contour
levels are: -6,-5,-4,-3,-2,-1, 1 mJy beam$^{-1}$.

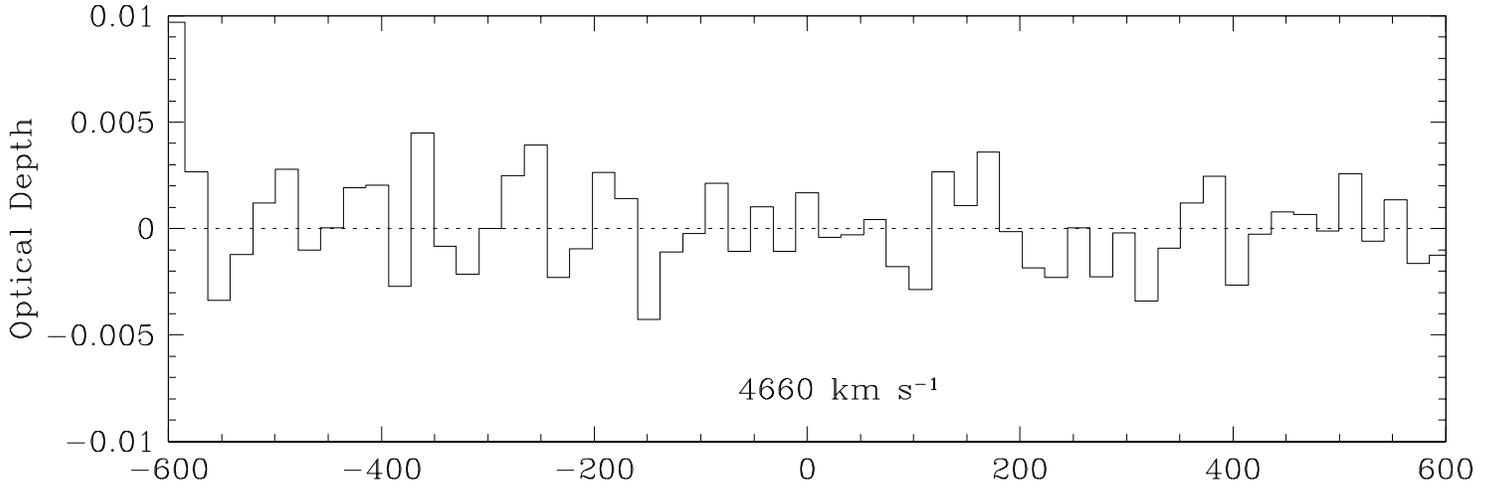
\noindent Figure 8 --  The HI 21cm absorption spectra of the Mrk 231
made with the VLBA on January 27, 1993 with a spatial resolution of
5$''$. All the spectra have been converted to optical depth using a
continuuum source flux density of 242 mJy. These spectra were made at
the velocities of the optical BAL systems seen toward Mrk 231.
The upper plot shows the spectrum centered at a
heliocentric redshift of z = 0.01554 (cz = 4660 km s$^{-1}$).
The middle plot shows the spectrum centered at a
heliocentric redshift of z = 0.02085 (cz =
6250 km s$^{-1}$). These two spectra have a spectral resolution of 
97.6 kHz. The lower plot shows the spectrum centered at a
heliocentric redshift of z = 0.02668 (cz = 8000 km s$^{-1}$), with 
a spectral  resolution of 390.6 kHz.

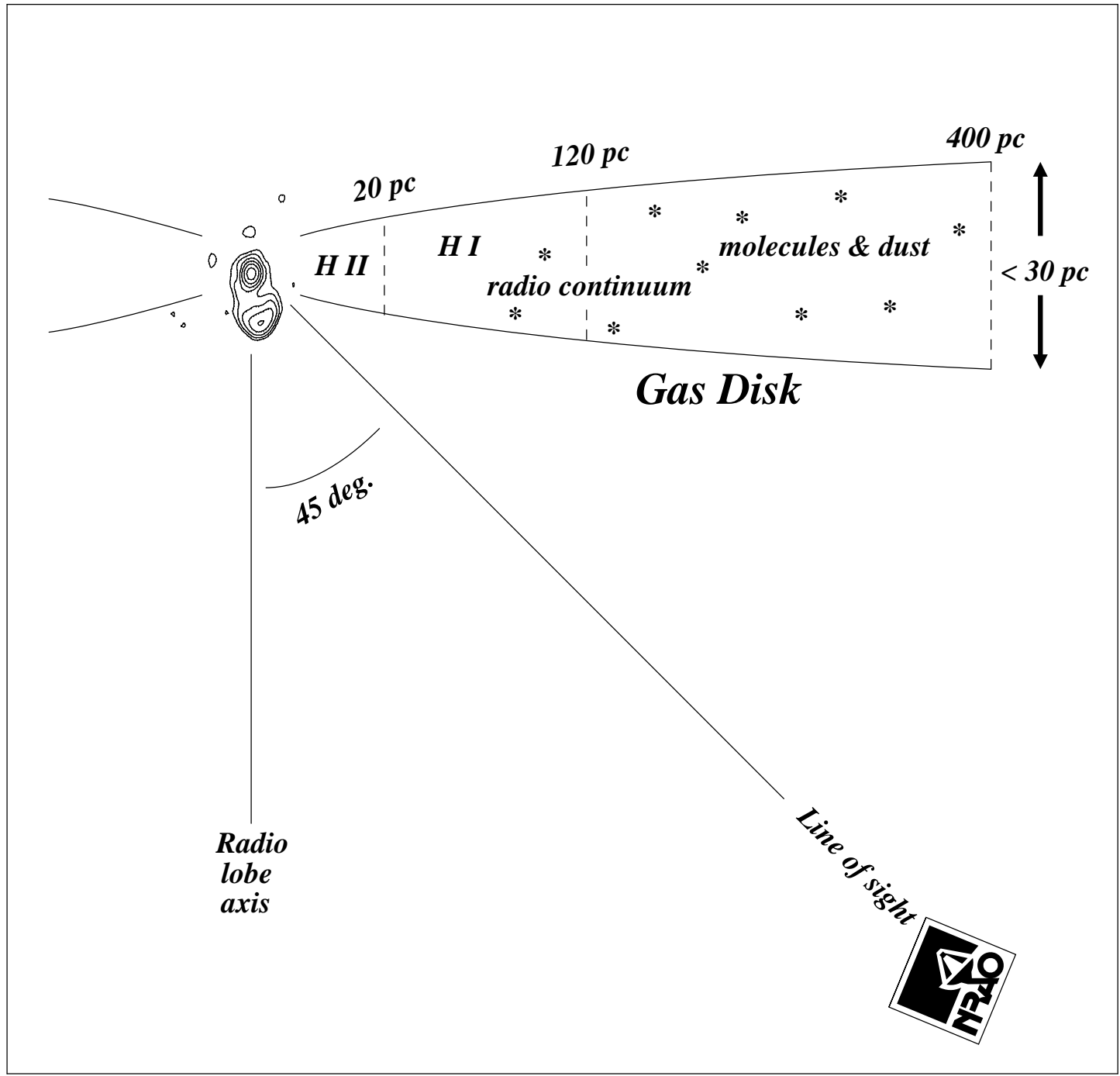
\noindent Figure 9 -- A schematic diagram of the model presented 
in sections 4.1 and 4.2  to explain the various observations of the
nuclear regions of Mrk 231.  The contours show the 
radio continuum emission from the pc-scale lobes and nucleus 
(Figure 2). The line drawing shows the multi-phase gaseous disk:
ionized atomic, neutral atomic, relativistic, and molecular gas and
dust. Note that the plot is not drawn strictly to scale, and that
the annotated values for size scales for the various gas phases 
are set mainly by observational limitations, ie. 
the data do not preclude a mixture  of the different
gaseous phases on  different scales.
The `stars' indicate possible
distributed massive star formation in the disk, and
the diffuse radio continuum emission from the disk is likely to be
driven by this star formation. In this model the
neutral atomic gas is roughly co-spatial  with the diffuse radio
continuum disk, thereby giving rise to the 
HI 21cm absorption seen by  the VLA and the shorter VLBI
baselines, while at the same time 
the radio nucleus and southern (approaching) pc-scale radio lobe
are not obscured by the HI disk.
 
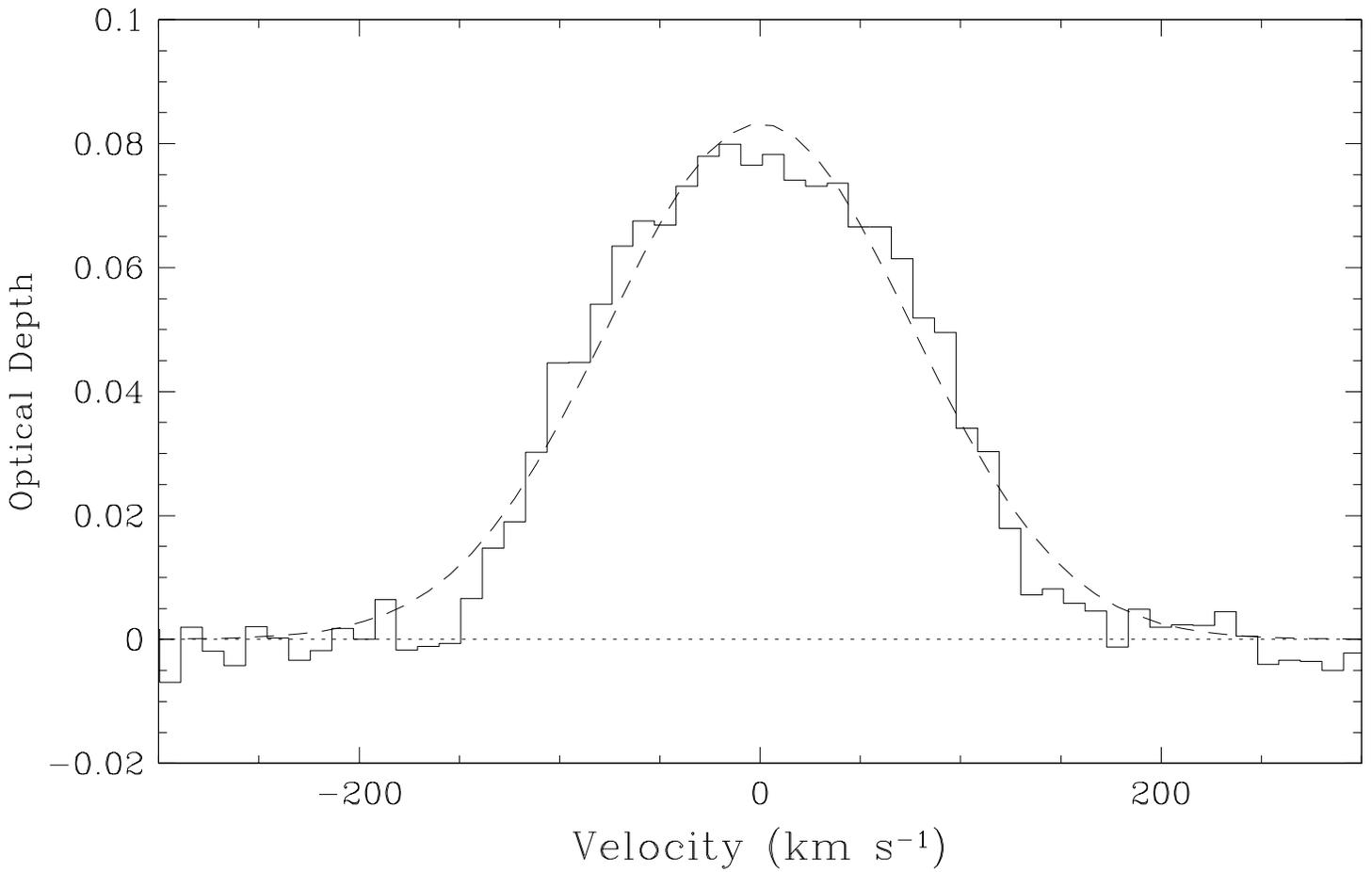
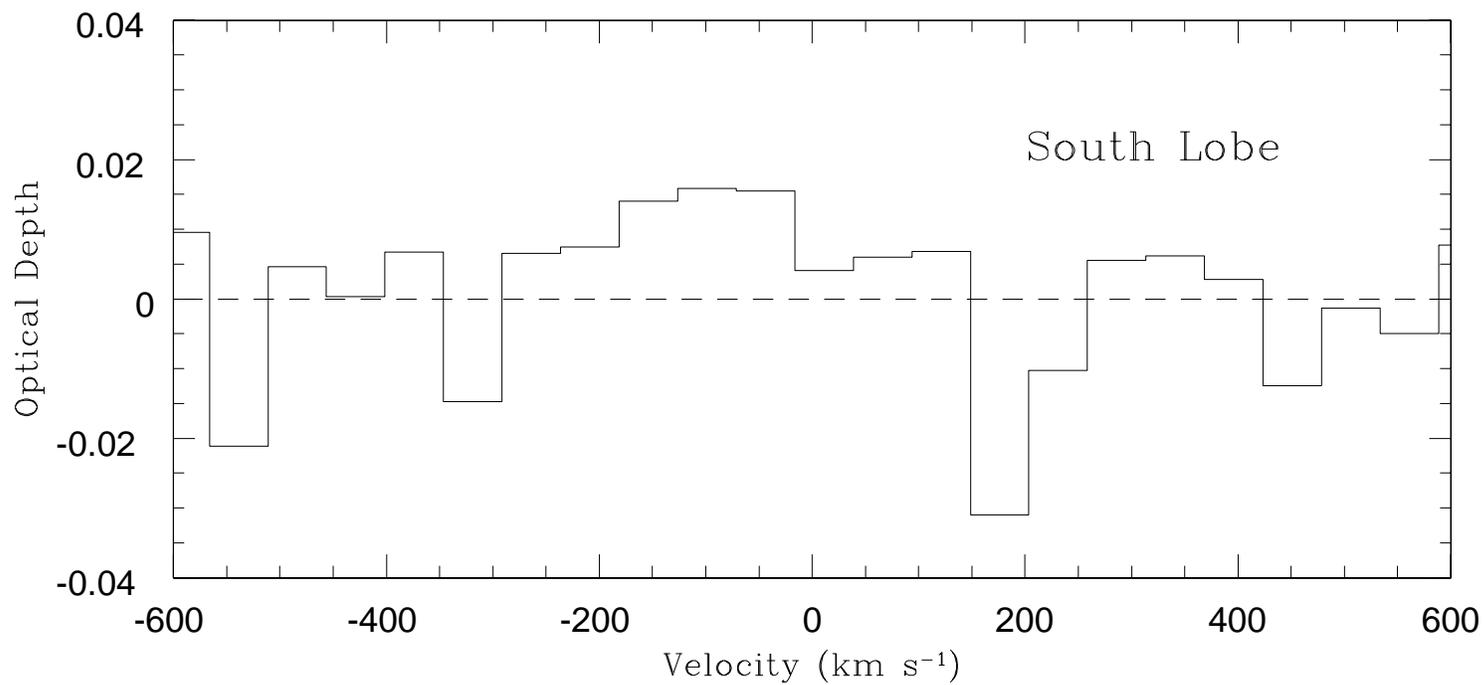
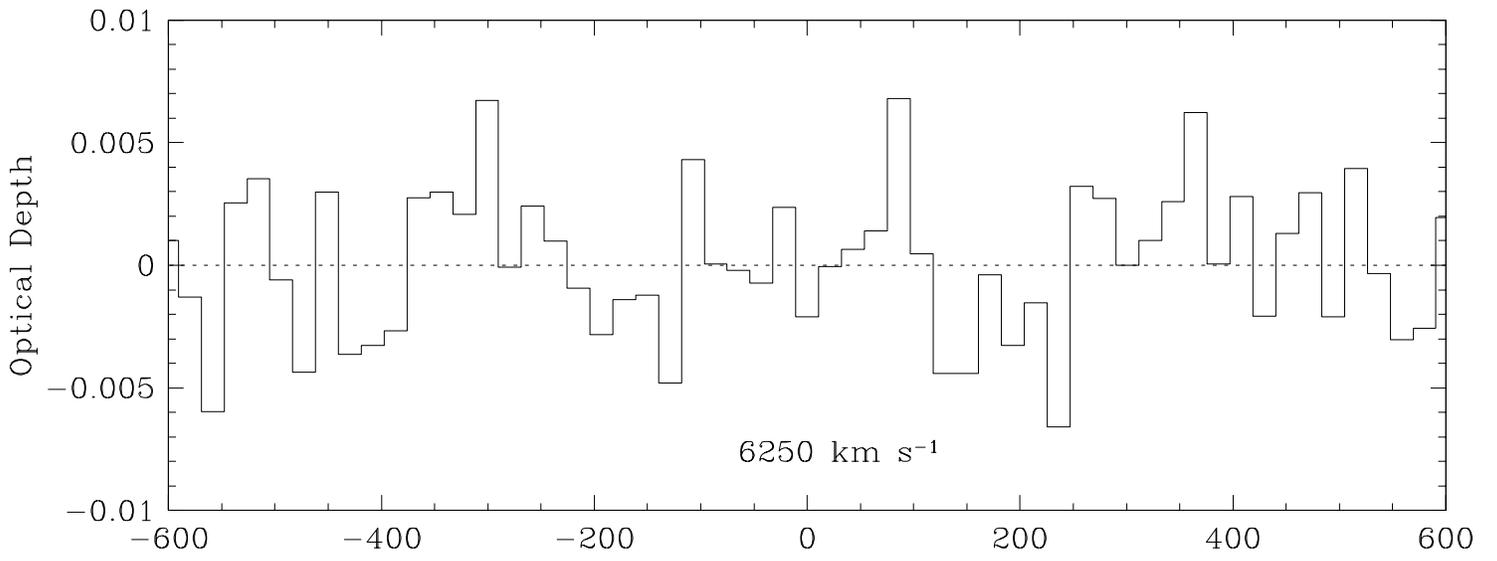
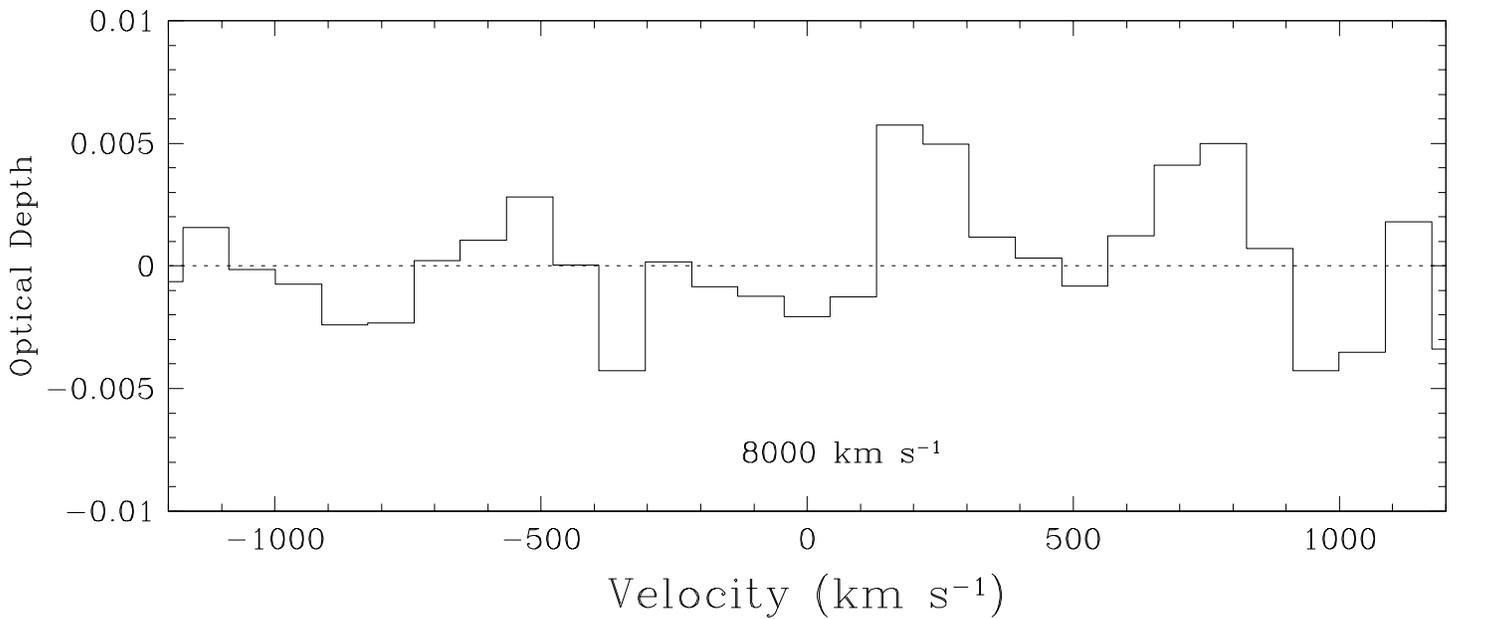
\end{document}